\documentclass[amsmath,amssymb,preprint,groupedaddress]{revtex4-1} 

\usepackage[english]{babel}
\usepackage[latin1]{inputenc} 

\begin{document}

\title{Equivalence between an extension of teleparallelism to a Weyl geometry and general relativity}
\author{J. B. Formiga}
\email[]{jansen.formiga@uespi.br}
\affiliation{Centro de Ciências da Natureza, Universidade Estadual do Piauí, C. Postal 381, 64002-150 Teresina, Piauí, Brazil}
\begin{abstract}
  Recently, an extension of teleparallelism to a Weyl geometry which allows us to easily establish conformal invariance and ``geometrize'' electromagnetism has been presented. In this paper, I extend a result which concerns the existence of the Schwarzschild solution  to a particular class of this extension. In addition, I obtain the field equations of some models based on this extension, including the one which is equivalent to Einstein's field equations with a massless scalar field.
\end{abstract}
\maketitle

\section{Introduction}
Since the advent of general relativity (GR), many new theories of gravity which attempt to generalize or give a different viewpoint of GR  have appeared in physics \cite{capozziello2011beyond,Hehlsdarticle1,o1997dawning,Sauer:2004hj}. Some of them are based on non-Riemannian geometries such as Riemann-Cartan and Weyl geometry. Among those theories, there is a particular class called teleparallelism \cite{Sauer:2004hj}, which corresponds to a very peculiar case of the Riemann-Cartan geometry (Weitzenböck spacetime) and has brought the attention of many physicists, including Einstein himself, who invented GR. 

Unlike GR, teleparallelism is based on the assumption that the curvature of the spacetime vanishes while the torsion tensor plays the role of gravity. These assumptions lead to the concept of distant parallelism, which is basically the existence of a rigid frame. Although, in some sense, this frame can be seen as a ``privileged frame'',  teleparallel models can be constructed in a frame independent way \cite{PhysRevD.67.044016}.

In the framework of teleparallelism, one can construct conformal theories and even a theory that is equivalent to GR, the so-called ``teleparallel equivalent of general relativity'' (TEGR). This means that we have a significant degree of freedom in this framework. Nonetheless, it was shown in Ref. \cite{PhysRevD.87.067702} that it is possible to take a richer approach by relaxing the teleparallel condition (distant parallelism) and assuming the existence of a peculiar nonmetricity tensor.

A very important result concerns the existence of a class of teleparallel models wider than TEGR that possesses the same solutions as GR does for a vast number of spacetimes, including all spacetimes that can be diagonalized by means of a coordinate change \cite{PhysRevD.88.068501}. In this paper, I extend this result to what I call ``Weyl teleparallel theory'' (WTT) and obtain  the field equations of a very general model. I also exhibit the WTT which is equivalent to Einstein's field equations with a massless scalar field.

The notation and conventions used throughout this paper is shown in Sec. \ref{662013c}, while a brief introduction to teleparallel theories is given in Sec. \ref{662013d}.  The results of this paper are all presented in Secs. \ref{662013b} and \ref{15102013a}, and  some final remarks are left to Sec. \ref{662013f}.

\section{Notation and conventions} \label{662013c}

Here I set the notation and conventions used throughout this paper. Let us start with the notation. The holonomic and anholonomic indices are denoted by Greek and Latin letters, respectively. The tetrad fields are represented by $e_a$ (frame) and $e^a$ (coframe), whose components in the coordinate basis are denoted by $e_a^{\  \mu}$ and $e^a_{\  \mu}$, respectively.  As usual, the coordinate basis is denoted by $\partial_{\mu}$, which stands for $\partial/\partial x^{\mu}$ with $x^{\mu}$ being a certain coordinate system. The components of the metric tensor in the tetrad basis are $\eta_{ab}=diag(+1,-1,-1,-1)$, while the ones in the coordinate basis are denoted by $g_{\mu \nu}$. I use square brackets around indices to represent the antisymmetric part of a tensor and round ones to indicate the symmetric part.

Let $M$ be a manifold endowed with a metric $g$,  a linear connection $\nabla$, and a Weyl $1$-form $\sigma$. Let $V,U,W$ be vectors belonging to the tangent bundle of $M$. The torsion tensor $T(V,U)$, the curvature $R(V,U)W$, and the Weyl $1$-form $\sigma$  are defined as follows:
\begin{eqnarray}
T(V,U)\equiv \nabla_V U-\nabla_U V-[V,U], \label{14012013a}\\
R(V,U)W\equiv \nabla_V\nabla_U W-\nabla_U\nabla_V W-\nabla_{[V,U]} W, \label{14012013b}\\
\sigma(V)g\equiv \nabla_V g. \label{2352013c}
\end{eqnarray}
Unless stated otherwise, the components of these tensors are defined as $T^a_{\ bc}\equiv <e^a, T(e_b,e_c)>$ and $R^a_{\  dbc}\equiv <e^a,R(e_b,e_c)e_d>$. In these definitions, I am using the standard convention $<df,\partial_{\mu}>=\partial_{\mu} f$, where $f$ is an arbitrary function.

The definition of nonmetricity and Weyl $1$-form vary a little bit in the literature. For instance, in Ref. \cite{Hehlsdarticle1}, the authors define them in such way that the nonmetricity $1$-form $Q_{ab}$ is related to $\sigma$ through the expression $Q_{ab}=-\eta_{ab}\sigma$, while their Weyl $1$-form, denoted by $Q$, is $Q=-\sigma$. Here, I have adapted their index notation to mine. One can easily check these relations by using equations (3.8.1) and (3.8.2), page 35, of Ref. \cite{Hehlsdarticle1}.

\section{Teleparallel theories} \label{662013d}
Teleparallel theories are based upon the assumption that the curvature of spacetime is zero and gravity is described by the torsion tensor. As a result, there exists an anholonomic frame $e_a$ in which the affine connection coefficients vanish, which is a peculiar feature of the Weitzenböck spacetime.  These assumptions can be written as
\begin{equation}
\nabla_{e_a}e_b=0 \label{2952013a}
\end{equation}
and
\begin{equation}
T(e_a,e_b)=-[e_a,e_b]. \label{2852013b}
\end{equation}

The Lagrangian of a ``general'' teleparallel model can be written in the form
\begin{equation}
T\equiv a_1T^{abc}T_{abc}+a_2T^{abc}T_{bac}+a_3T^aT_a, \label{2852013a}
\end{equation}
where $T_a\equiv T^b_{\  ba}$.

From Eq. (\ref{2852013b}), we obtain
\begin{equation}
T^a_{\  bc}=2e_b^{\  \mu}e_c^{\  \nu} e^a_{ \ [\nu,\, \mu]}. \label{2852013c}
\end{equation}

It is well known that the  Lagrangian
\begin{equation}
\mathring{T}\equiv -\frac{1}{4}T^{abc}T_{abc}-\frac{1}{2}T^{abc}T_{bac}+T^aT_a, \label{2852013h}
\end{equation}
which corresponds to the TEGR model, is equivalently to that of GR. Therefore, the choice  $a_1=-1/4$, $a_2=-1/2$, and $a_3=1$ corresponds to a very special case of $T$.

Since the field equations of both TEGR and GR are the same, they have the same solutions in vacuum. As a result, the TEGR agrees with all experiments performed so far.

\section{Weyl Teleparallel Theory} \label{662013b}
It was pointed out in Ref. \cite{PhysRevD.88.068501} that the teleparallel model characterized by $a_3=1$ and  $a_2=-1-2a_1$, with $a_1$ arbitrary, has the same vacuum field equations  as GR does for any spacetime that can be diagonalized by a coordinate change. Here I extend this result  to the WTTs. But first, let us see some fundamental features of these models ( for more details, see Ref. \cite{PhysRevD.87.067702}).

The WTT is probably the simplest generalization of teleparallelism to a Weyl geometry. It is not a teleparallel theory because, for a non-integrable Weyl geometry, there is no frame that satisfies (\ref{2952013a}). Nonetheless, there is a frame $e_a$ which satisfies
\begin{equation}
\nabla_{e_b}e_a=-\frac{1}{2} \sigma_b e_a, \label{2952013b}
\end{equation}
where $\nabla$ is the affine connection, and $\sigma$ is the Weyl $1$-form \footnote{In terms of the connection $1$-form $\omega_{ab}$, defined as $\omega_{ab}=\eta_{ad}<e^d,\nabla_{e_c}e_b> e^c$, Eq. (\ref{2952013b}) can be expressed as $\omega_{ab}=-\sigma\eta_{ab}/2$ or, in terms of the Weyl $1$-form defined in Ref. \cite{Hehlsdarticle1}, as $\omega_{ab}=Q\eta_{ab}/2$.}.

In this kind of theory, we can represent Weyl transformations through the changes
\begin{equation}
\tilde{e}_a=e^{-\theta}e_a,\quad \tilde{\sigma}=\sigma+2d\theta, \label{2952013c}
\end{equation}
where  $\theta$ is an arbitrary scalar function, $d$ is the exterior derivative operator, and the tilde represents quantities in another Weyl frame. It is easy to verify that Eq. (\ref{2952013b}) is invariant under the transformations (\ref{2952013c}). However, the WTT models considered here are not invariant under these transformations. 

As shown in Ref. \cite{PhysRevD.87.067702}, the Lagrangian
\begin{equation}
\overset{\bullet}{T}=\biggl(-\frac{1}{4}\hat{T}^{abc}\hat{T}_{abc}-\frac{1}{2}\hat{T}^{abc}\hat{T}_{bac} +\hat{T}^a\hat{T}_a+\frac{3}{2}\sigma_a\sigma^a-2\sigma_a\hat{T}^a \biggr) \label{262013a}
\end{equation}
is equivalent to that of GR, where $\hat{T}^{a}_{ \ bc}$ is the torsion components. In the Weyl frame $(e_a,\sigma)$, these components take the form
\begin{equation}
\hat{T}^a_{  \ bc}=2e_b^{ \ \mu}e_c^{ \ \nu} e^a_{ \ [\nu,\, \mu]} +\sigma_{[c|}\delta^a_{ \ |b]}. \label{3152013e}
\end{equation}
It follows that
\begin{eqnarray}
\hat{T}^a\hat{T}_a=4e_a^{ \ \mu} e_b^{ \ \beta} g^{\alpha \nu} e^a_{ \ [\nu,\mu]} e^b_{ \ [\alpha,\beta]}+6e_a^{ \ \mu}\sigma^{\nu}e^a_{ \ [\nu,\mu]}
\nonumber \\
+(9/4)\sigma_a\sigma^a,
\label{3152013a} \\
\hat{T}^{abc}\hat{T}_{abc}=4g^{\mu \beta} g^{\alpha \nu}e^a_{ \ [\nu,\mu]} e_{a [\alpha,\beta]}+4e_a^{ \ \mu}\sigma^{\nu}e^a_{ \ [\nu,\mu]}
\nonumber \\
+(3/2)\sigma_a\sigma^a,
\label{3152013b} \\
\hat{T}^{abc}\hat{T}_{bac}=4e_b^{ \ \mu} e_a^{ \ \beta} g^{\alpha \nu} e^a_{ \ [\nu,\mu]} e^b_{ \ [\alpha,\beta]}+2e_a^{ \ \mu}\sigma^{\nu}e^a_{ \ [\nu,\mu]}
\nonumber \\
+(3/4)\sigma_a\sigma^a.
\label{3152013c}
\end{eqnarray}

One can generalize the Lagrangian $\overset{\bullet}{T}$ by taking the following arbitrary combination
\begin{eqnarray}
\hat{T}=a_1\hat{T}^{abc}\hat{T}_{abc}+a_2\hat{T}^{abc}\hat{T}_{bac} +a_3\hat{T}^a\hat{T}_a+a_4\sigma_a\sigma^a
\nonumber \\
+a_5\sigma_a\hat{T}^a. \label{262013b}
\end{eqnarray}

\subsection{Diagonalizable metric} \label{30072013f}

From Eqs. (\ref{3152013b}) and (\ref{3152013c}), we see that the terms with $\sigma$ in $\hat{T}^{abc}\hat{T}_{abc}-2\hat{T}^{abc}\hat{T}_{bac}$  cancel out independently of the tetrad field. As we shall see next, this fact allows us to ensure the equivalence between the Lagrangians $\hat{T}$ and $\overset{\bullet}{T}$ for a diagonalizable metric without imposing any restriction on $\sigma$.

By imposing  $a_3=1$, $a_2=-1-2a_1$, and $\hat{T}=\overset{\bullet}{T}$, we arrive at
\begin{equation}
T^{abc}T_{abc}-2T^{abc}T_{bac} +\frac{a_4-3/2}{1/4+a_1}\sigma_a\sigma^a+\frac{a_5+2}{1/4+a_1}\sigma_a\hat{T}^a=0, \label{3152013d}
\end{equation}
where I have already assumed that $a_1 \neq -1/4$ and used $\hat{T}^{abc}\hat{T}_{abc}-2\hat{T}^{abc}\hat{T}_{bac}=T^{abc}T_{abc}-2T^{abc}T_{bac}$. Remember that $T^a_{\ bc}$ is the Weitzenböck torsion, which is given by Eq. (\ref{2852013c}).

If we assume that the metric is diagonalizable, then  $T^{abc}T_{abc}=2T^{abc}T_{bac}$ \cite{PhysRevD.88.068501}. In this case,  Eq. (\ref{3152013d}) reduces to
\begin{equation}
(a_4-3/2)\sigma_a\sigma^a+(a_5+2)\sigma_a\hat{T}^a=0. \label{2352013a}
\end{equation}
From Eq. (\ref{2352013a}), we see that it is natural to choose $a_4=3/2$ and $a_5=-2$. In this case, the Lagrangian $\hat{T}$ will yield the same field equations as $\overset{\bullet}{T}$ does for diagonalizable spacetimes. This means that, for this kind of spacetime, the model (\ref{262013b}) with $a_3=1$, $a_2=-1-2a_1$, $a_4=3/2$, and $a_5=-2$,  possesses the same solutions as GR. The models with these parameters can be considered a generalization of the so called ``new general relativity'' \cite{PhysRevD.19.3524}.

Let us now analyze the possibility of other combinations of $a_4$ and $a_5$ that satisfies Eq. (\ref{2352013a}). From Eq. (\ref{3152013e}), we see that 
\begin{equation}
\sigma_c T^c=2\sigma^{\nu}e_{a}^{ \ \mu} e^a_{ \ [\nu,\mu]}+\frac{3}{2} \sigma_c\sigma^c. \label{2352013b}
\end{equation}
By comparing Eq. (\ref{2352013b}) with (\ref{2352013a}) and considering $\sigma$ as independent of the tetrad, we may conclude that there will not be any other choice for $a_4$ and $a_5$, unless the first term on the right-hand side of Eq. (\ref{2352013b}) vanishes. One can easily check that this term is not zero for an arbitrary diagonal metric.

\section{Field equations} \label{15102013a}
The Lagrangian (\ref{262013b}) can be split into two parts
\begin{equation}
\hat{T}=T+W, \label{162013a}
\end{equation}
where

\begin{eqnarray}
T=4\Biggl( a_1g^{\mu \beta}\eta_{ab}+a_2e_a^{ \ \beta}e_b^{ \ \mu}
\nonumber \\
+a_3e_a^{ \ \mu}e_b^{ \ \beta}   \Biggr) g^{\alpha \nu} e^a_{ \ [\nu,\mu]} e^b_{ \ [\alpha,\beta]}, \label{662013i}
\\
W=2Ae_a^{ \ \mu} \sigma^{\nu} e^a_{ \ [\nu,\mu]}
+(3/4) B\sigma_a\sigma^a, \label{662013o}
\end{eqnarray}
with
\begin{equation} 
A\equiv 2a_1+a_2+3a_3+a_5, \quad B\equiv 2a_1+a_2+3a_3+4a_4/3+2a_5. \label{662013m}
\end{equation}
Note that $W$ accounts for all terms with $\sigma$. Note also that the constants $A$ and $B$ vanish for $a_3=1$, $a_2=-1-2a_1$, $a_4=3/2$, and $a_5=-2$.

From the Lagrangian (\ref{162013a}), we can write the field equations obtained from the variations with respect to the metric in the form
\begin{equation}
\hat{\mathbb{T}}^{\theta \psi}\equiv \frac{1}{e}\frac{\delta e \hat{T}}{\delta g_{\theta \psi}}=\mathbb{T}^{\theta \psi}+\mathbb{W}^{\theta \psi}, \label{162013c}
\end{equation}
where $\mathbb{W}^{\theta \psi}$ is the part that accounts for the Weyl field, and $e$ is the determinant of the tetrad field.

\subsection{Field equations for independent variations of $g_{\mu \nu}$ and $\sigma_{\mu}$}
Treating the variations $\delta g_{\mu \nu}$ and $\delta \sigma_{\mu}$ as independent, we get 

\begin{equation}
\mathbb{T}^{\theta \psi}=\frac{1}{2}[ g^{\theta \psi}T+e^{c(\psi|} I_c^{ \ |\theta)}-e_a^{ \ \lambda}e^a_{ \ \lambda, \sigma} e^{c(\psi|} I_c^{ \ |\theta)\sigma}-e^{c(\psi|} I_{c \ \quad,\sigma}^{ \ |\theta)\sigma}    ] \label{30072013b}
\end{equation}
and 
\begin{equation}
\mathbb{W}^{\theta \psi}=\frac{1}{2}[ g^{\theta \psi}W+e^{c(\psi|} W_c^{ \ |\theta)}-e_a^{ \ \lambda}e^a_{ \ \lambda, \sigma} e^{c(\psi|} W_c^{ \ |\theta)\sigma}-e^{c(\psi|} W_{c \ \quad,\sigma}^{ \ |\theta)\sigma}    ], \label{30072013d}
\end{equation}
where
\begin{eqnarray}
I_c^{ \ \lambda}= -8\Biggl[ 2a_1e_c^{ \ (\mu|} g^{\lambda |\beta)}g^{\alpha \nu}\eta_{ab}
\nonumber \\
+\left(a_2e_a^{ \ \beta}e_b^{ \ \lambda}+a_3e_a^{ \ \lambda}e_b^{ \ \beta} \right)e_c^{ \ \mu} g^{\alpha \nu}
\nonumber \\
+\left(a_2e_a^{ \ \beta}e_b^{ \ \mu}+a_3e_a^{ \ \mu}e_b^{ \ \beta} \right) e_c^{ \ (\alpha|} g^{\lambda |\nu)} \Biggr] e^a_{ \ [\nu,\mu]} e^b_{ \ [\alpha,\beta]}, \label{662013j}
\end{eqnarray}

\begin{eqnarray}
I_c^{ \ \lambda \sigma}= 8\Biggl(a_1g^{\nu[\lambda|}g^{\mu |\sigma]} \eta_{ac}+a_2e_a^{ \ [\sigma|} g^{\nu |\lambda]}e_c^{ \ \mu}
\nonumber \\
+a_3e_c^{ \ [\sigma|} g^{\nu |\lambda]}e_a^{ \ \mu} \Biggr)e^a_{ \ [\nu,\mu]}, \label{662013l}
\end{eqnarray}

\begin{eqnarray}
W_c^{ \ \lambda}= 2AL_{ca }^{\quad \ \lambda \nu \mu \alpha} \sigma_{\alpha} e^a_{ \ [\nu,\mu]}
 -(3/2)B\sigma_c \sigma^{\lambda}, \label{162013b}  
\end{eqnarray}

\begin{equation}
W_c^{ \ \lambda \sigma}= 2Ae_c^{  \ [\sigma } \sigma^{\lambda]}. \label{862013a}
\end{equation}
The term $L_{ca }^{\ \ \lambda \nu \mu \alpha}$ in Eq. (\ref{162013b}) has been defined as
\begin{equation}
L_{ca }^{\ \ \lambda \nu \mu \alpha}\equiv e_a^{ \ [\nu|} g^{\lambda |\mu]}e_c^{ \ \alpha}+e_c^{ \ [\nu|} g^{\alpha |\mu]}e_a^{ \ \lambda}+e_a^{ \ [\nu|} e_c^{ \ |\mu]} g^{\lambda \alpha}. \label{11102013a}
\end{equation}

The objects (\ref{662013j})--(\ref{862013a}) have been defined as $I_c^{ \ \lambda}\equiv \partial T/\partial e^c_{  \ \lambda}$,  $W_c^{ \ \lambda}\equiv \partial W/\partial e^c_{  \ \lambda}$, $I_c^{ \ \lambda \sigma}\equiv \partial T/\partial e^c_{  \ \lambda,\sigma}$ and $W_c^{ \ \lambda \sigma}\equiv \partial W/\partial e^c_{  \ \lambda,\sigma}$. From the expressions (\ref{662013j})--(\ref{11102013a}), it is clear that these objects can be regarded as tensor components if we use Eq. (\ref{3152013e}) to rewrite $e^a_{ \ [\nu, \mu]}$ in terms of $\hat{T}^a_{ \ \mu \nu}$ and $\sigma$. Of course, in an arbitrary tetrad field both $\mathbb{T}^{\theta \psi}$ and $\mathbb{W}^{\theta \psi}$ will depend on $\sigma$.

By varying $e\hat{T}$ with respect to $\sigma_{\alpha}$, one obtains
\begin{equation}
W^{\alpha}\equiv \frac{\delta \hat{T}}{\delta \sigma_{\alpha}}=2Ae_a^{ \ \mu} g^{\alpha \nu}e^a_{ \ [\nu,\mu]}+\frac{3}{2}B\sigma^{\alpha}, \label{662013n}
\end{equation}
which can be rewritten in the form
\begin{equation}
W^{\alpha}=A\hat{T}^{\alpha}+\frac{3}{2} \left( \frac{4}{3}a_4+a_5\right)\sigma^{\alpha} \label{30072013a}.
\end{equation}

The field equations of teleparallel theories can be obtained by setting $\sigma=0$ , while Einstein's field equations can be obtained by setting $a_1=-1/4$, $a_2=-1/2$, $a_3=1$, $a_4=3/2$, and $a_5=-2$. The field equation (\ref{162013c}) is also equivalent to Einstein's one for the values $a_3=1$, $a_2=-1-2a_1$, $a_4=3/2$, and $a_5=-2$ when $g$ is diagonalizable. These results were checked using Eqs. (\ref{30072013b})--(\ref{11102013a}).

It was shown in Ref. \cite{PhysRevD.87.067702} that when we add the term $4R^{\mu \nu}R_{\mu \nu}$, where $R_{\mu \nu}=R^{\lambda}_{\ \mu \lambda \nu}$, to the Lagrangian (\ref{262013a}) and identify $\sigma$ with twice the electromagnetic potential,  we obtain Einstein's field equations with the Maxwell energy-momentum tensor. Putting this together with the result of Sec. \ref{30072013f}, we conclude that if we add the term $4R^{\mu \nu}R_{\mu \nu}$ to the Lagrangian (\ref{262013b}) and take $a_1=-1/4$, $a_2=-1-2a_1$, $a_3=1$, $a_4=3/2$, and $a_5=-2$, we will have a model with  the same solutions as GR does in the presence of electromagnetic field whenever the spacetime can be diagonalized.

\subsection{ The WTT model that is equivalent to a massless scalar field in GR}
Now I construct a WTT model which yields the field equations of GR coupled with a massless scalar field. To achieve that, I  consider an integrable Weyl geometry and regard a scalar field $\varphi$ as independent of $g_{\mu \nu}$. In this case, the Weyl $1$-form is assumed to be given by $\sigma_{\mu}=\varphi_{,\mu}$. It is clear, then, that $\sigma$ remains independent of $g_{\mu \nu}$, although the variations now are taken with respect to $g$ and $\varphi$.

In this new approach, Eq. (\ref{162013c}) remains the same. However, Eq. (\ref{30072013a}) changes when we perform the variation of $e\hat{T}$ with respect to $\varphi$ rather than $\sigma_{\mu}$. 

If we assume that $A$ vanishes and $a_1=-1/4$, $a_2=-1/2$, and $a_3=1$, we will have $a_5=-2$; nonetheless, the parameter $a_4$ remains arbitrary. These assumptions implies that the tensor (\ref{30072013b}) equals minus the Einstein one. From Eqs. (\ref{662013o}), (\ref{30072013b}), (\ref{30072013d}), (\ref{162013b}) and (\ref{862013a}),   we see that Eq. (\ref{162013c}) can be rewritten as
\begin{equation}
G^{\mu \nu}=-\frac{3}{4}B\left( \sigma^{\mu}\sigma^{\nu}-\frac{1}{2}g^{\mu \nu}\sigma_{\mu}\sigma^{\mu}  \right), \label{30072013c}
\end{equation}
where $G^{\mu \nu}$ is the Einstein tensor in terms of the Levi-Civita symbols. Note that $B$ is not necessarily zero, since $a_4$ remains arbitrary.

The variation of (\ref{162013a}) with respect to $\varphi$ yields
\begin{equation}
\Box \varphi=0, \label{30072013e}
\end{equation}
where $\Box$ is the d'Alembertian operator, which is also written in terms of the Levi-Civita symbols.

The field equations (\ref{30072013c}) and (\ref{30072013e}) are the Einstein field equations with a massless scalar field. Their spherically symmetric solution is well known in the literature \cite{PhysRevD.24.839}. Furthermore, it has been shown that the presence of the scalar field does not change the solar-system experiments for a large range of values of the coupling constant\footnote{The coupling constant is proportional to $B$ and can be related to the ``scalar charge'' in Ref. \cite{PhysRevD.83.087502}.} \cite{PhysRevD.83.087502}.

\section{Final Remarks} \label{662013f}

With respect to the coupling of matter fields, it has been shown that the TEGR cannot couple consistently with the Dirac spinor \cite{PhysRevD.67.044016}, unless one abandons the concept of absolute parallelism for spinors by using the Riemannian connection when coupling a spinor field with gravity \cite{PhysRevD.67.108501}. Nonetheless, this inconsistency is not present in the models based on the values $a_3=1$, $a_2=-1-2a_1$, $a_1 \neq -1/4$ [see, e.g., p. 6 of Ref. \cite{PhysRevD.67.044016} and Eq. (5) of Ref. \cite{PhysRevD.88.068501}].

Perhaps the most interesting aspect of the WTTs is the facility to create theories which are invariant under the transformations (\ref{2952013c}). Besides, as pointed out in Ref. \cite{PhysRevD.87.067702},  some versions of the WTTs may be free from the ``second clock effect'', even in the case of a non-integrable Weyl geometry.


%

\end{document}